\documentstyle[11pt,iopconf]{article}

\newcommand{\gton}{\stackrel{\large >}{\sim}}
\newcommand{\beq}{\begin{equation}}
\newcommand{\beqar}{\begin{eqnarray}}
\newcommand{\eeq}[1]{\label{#1} \end{equation}}
\newcommand{\eeqar}[1]{\label{#1} \end{eqnarray}}

\begin{document}
\renewcommand{\bottomfraction}{0.95}
\renewcommand{\topfraction}{0.95}
\renewcommand{\textfraction}{0.05}
\title{Applications of a Global Nuclear-Structure Model to Studies of
the Heaviest Elements}
\author{Peter M\"{o}ller\dag\ and J. Rayford Nix\dag}
\affil{\dag\ Theoretical Division,
LANL, Los Alamos, NM 87545, USA}
\beginabstract
We present some new results on heavy-element nuclear-structure properties
calculated on the basis of the finite-range droplet model and folded-Yukawa
single-particle potential.  Specifically, we discuss calculations of nuclear
ground-state masses and microscopic corrections, $\alpha$-decay properties,
$\beta$-decay properties, fission potential-energy surfaces, and
spontaneous-fission half-lives.  These results, obtained in a global
nuclear-structure approach, are particularly reliable for describing the
stability properties of the heaviest elements.
\endabstract
\baselineskip=18.2pt plus 0.2pt minus 0.2pt
\lineskip=18.2pt plus 0.2pt minus 0.2pt

\section{Introduction}

The number of elements is limited because nuclei become
increasingly unstable
with respect to spontaneous fission and $\alpha$ decay
as the proton number increases.
\begin{figure}[t]
\vspace{3.8in}
\caption[masd92a]{\baselineskip=12pt\small
Comparison of experimental  and calculated microscopic corrections for
1654 nuclei, for a macroscopic model corresponding to the finite-range
droplet model. The bottom part showing the difference between these two
quantities is equivalent to the difference between measured and
calculated ground-state masses. There are almost no systematic errors
remaining for nuclei above $N=65$, for which region the error is only
0.448  MeV\@.
}
\label{masd92a}
\end{figure}
\begin{figure}[t]
\vspace{3.8in}
\caption[comp2adj]{\baselineskip=12pt\small
Test of extrapability of the FRDM
towards the superheavy region. The top part of the figure shows
the error of the standard FRDM\@. In the lower part the error was
obtained from a mass model whose constants were determined from
adjustments to the restricted set of nuclei with $Z,N \ge 28$ and $A
\le 208$. In the
heavy region there is some increase in the spread of the error, but no
systematic divergence of the mean error.
}
\label{comp2adj}
\end{figure}
\begin{figure}[t]
\vspace{3.5in}
\caption[actmic]{\baselineskip=12pt\small
Contour diagram of calculated microscopic corrections at the end of the
periodic system.  Solid squares indicate nuclei that are calculated to
be stable with respect to $\beta$ decay.  The
well-known doubly magic nucleus $^{208}_{\phantom{0}82}$Pb$_{126}$ is
associated with the minimum in the shell correction in the
lower-left-hand corner. Superheavy nuclei are associated with the
minimum in the upper-right-hand corner.}
\label{actmic}
\end{figure}
\begin{figure}[t]
\vspace{3.5in}
\caption[qcomp]{\baselineskip=12pt\small
Comparison of measured and calculated $\alpha$-decay $Q$ values for the
$N=154$ and 155 isotonic chains.
}
\label{qcomp}
\end{figure}
\begin{figure}[t]
\vspace{7.0in}
\caption[hulet]{\baselineskip=12pt\small
Experimental fission-fragment mass and kinetic-energy distributions for
the fission  of nuclei close to $^{264}$Fm, whose symmetric fragments
are doubly magic. The structures of these distributions reflect the
valleys, ridges, minima, and saddle points of the underlying nuclear
potential-energy surfaces.
}
\label{hulet}
\end{figure}
\begin{figure}[t]
\vspace{3.5in}
\caption[shaps]{\baselineskip=12pt\small
Nuclear shapes for which fission potential-energy surfaces are
calculated. The selected shapes allow fission into both compact
spherical fragments with high kinetic energies and elongated fragments
with normal kinetic energies.
}
\label{shaps}
\end{figure}
\begin{figure}[t]
\vspace{3.5in}
\caption[fm258]{\baselineskip=12pt\small
Calculated potential-energy surface for $^{258}$Fm, showing three paths
to fission. Initially, only one path starting at the ground state
exists. Later this path divides into two paths, one leading to compact
scission shapes in the lower part of the figure and the other leading
to elongated shapes in the upper part of the figure. At a late stage in
the barrier-penetration process, a third ``switchback path'' branches
off from the path leading to compact shapes and leads back into the
valley of elongated scission shapes. Because this takes place late in
the barrier-penetration process, the fission probabilities for fission
into compact and elongated shapes are expected to be roughly
comparable.  Experimentally the probabilities differ by only one order
of magnitude. The inertia associated with fission into the lower valley
is considerably smaller than the inertia for fission into the upper
valley.
}
\label{fm258}
\end{figure}
\begin{figure}[t]
\vspace{3.5in}
\caption[shapa]{\baselineskip=12pt\small
Shapes corresponding to the contour map in fig.~\ref{cona258}.  Shapes
associated with the new valley are in the lower part of the figure and
remain symmetric. As the switchback path from the new valley crosses
over the saddle at $r=1.4$, $\sigma=0.75$ into the old valley,
asymmetry becomes more and more pronounced. As asymmetry develops, the
overall extension of the nucleus remains approximately constant for
fixed values of $r$.
}
\label{shapa}
\end{figure}
\begin{figure}[t]
\vspace{2.8in}
\caption[cona258]{\baselineskip=12pt\small
Contour map for $^{258}$Fm, showing the vicinity of the outer saddle
along the new valley and the saddle along the switchback path between
the new valley and the old valley.  The energy has been minimized with
respect to the mass-asymmetry coordinate $\alpha_2$ for fixed values of
the other symmetric three-quadratic-surface shape parameters.
The new valley enters in the extreme lower left of this
figure, and fission may either evolve into the old valley across the
saddle at $r=1.4$, $\sigma=0.75$ or proceed in the direction of compact
scission shapes across the saddle at $r=1.6$, $\sigma=0.74$.  These two
saddles are of about equal height.}
\label{cona258}
\end{figure}
\begin{figure}[t]
\vspace{4.5in}
\caption[fhalf]{\baselineskip=12pt\small
Experimental spontaneous-fission half-lives compared to calculated values
for fission along the old and new valleys. A new valley is present
in the calculated potential-energy surface only for $N\geq158$.
When half-lives have been calculated for both valleys for a particular
neutron number, the shorter (dominating)
calculated half-lives should be compared with experimental values.
The discrepancy between calculated and experimental values in the
vicinity of $N=152$ may arise from either an error
in the calculated ground-state
energy or the neglect of fission along the third (switchback)
path.

For No there is a new experimental feature of fairly constant half-life
for $N$\hspace{0.5ex}\raisebox{-0.6ex}{$\gton$}\hspace{0.5ex}$156$,
which is  reproduced moderately well by the
calculations.

For Rf the
experimental half-live is nearly constant  as a function of
$N$. The theoretical half-lives for Rf are too high near $N=152$.
However, the discrepancy corresponds only to an error of about 1 MeV
in the calculated ground-state energy.

For $Z=106$ the  calculated half-life in the new valley is fairly
constant beyond $N=156$. This shows that the destabilizing effect of
the spherical magic-fragment neutron number $N= 2\times 82$
approximately cancels the effect of the deformed magic-ground-state
neutron number $N=162$.
}
\label{fhalf}
\end{figure}
Already in the mid-1960's it was speculated that this
trend might be broken at the next magic numbers beyond those in the
doubly magic nucleus $^{208}_{\phantom{0}82}$Pb$_{126}$.
Many calculations on the properties of the heaviest elements were
carried out over the next several years.
However, since that time
significant improvements have been incorporated into the model that we
use for these studies and we present here some of our most recent results.
More extensive presentations will appear in a forthcoming
review [1] and
in a forthcoming issue of {\it Atomic Data and Nuclear Data
Tables\/} [2]. The new results are
particularly reliable in the heavy-element region.

\section{Model}

In the macroscopic-microscopic method the
total potential energy, which is calculated as
a function of shape, proton number $Z$\/, and neutron number $N$\/, is the
sum of a macroscopic term and a microscopic term
representing the shell-plus-pairing correction.
Thus, the total nuclear
potential energy can be written as
\beq
E_{\rm pot}(Z,N,{\rm shape})= E_{\rm mac}(Z,N,{\rm shape})+
E_{\rm s+p}(Z,N,{\rm shape})
\eeq{toten}
The preferred model in the current calculations has its origin in a 1981
nuclear mass model [3,4], which utilized the
folded-Yukawa single-particle potential developed in
1972 [5,6].  The macroscopic model used in the
1981 calculation was a finite-range liquid-drop model, which contained a
modified surface-energy term to account for the finite range of the
nuclear force.  The modified surface-energy term was given by the
Yukawa-plus-exponential finite-range model~[7].
This model is used in our calculation of fission potential-energy surfaces.

Our preferred macroscopic model is now the
finite-range droplet model, for which
additions of finite-range surface-energy effects and an exponential
term [8] have resulted in dramatic
improvements in its predictive properties, as summarized in the
discussion of Table A in Ref.~[9]. We refer to this new
macroscopic model as the finite-range droplet-model (FRDM), which abbreviation
is also used to designate the full macroscopic-microscopic nuclear-structure
model. For the calculation of ground-state properties we use here
the latest version, which is denoted by FRDM (1992)~[2].

\section{Ground-state properties}

Figure \ref{masd92a} shows the results of the FRDM (1992) nuclear-mass
calculation. The discrepancy between measured and calculated masses
shown in the lower part of the figure is quite small, in
particular in the heavy region.
The good agreement results from several essential new features in the
calculation relative to
those in the 1981 calculation [3,4],
namely a new macroscopic model, a Lipkin-Nogami pairing model with an
improved form and parameters of the effective-interaction pairing
gap [10],
and minimization of the ground-state energy with respect to higher-multipole
shape distortions [11]. The FRDM accounts for Coulomb
redistribution effects, which are particularly important in
the heavy region [11].
To assure the reliability of a
nuclear mass model for extrapolation to the superheavy region, it is
in our view necessary to use
a {\it global\/} approach in which the model constants
are adjusted
to a large region of the
periodic system, as is done here.
Approaches in which the model
constants are adjusted to a limited heavy region,
such as the region above
Pb, are much less reliable for extrapolation into the superheavy region.

To test the reliability of the FRDM for extrapolation beyond the
heaviest known elements we have performed a rather severe test
in which we
adjust the model constants only to data in the region $Z,N \ge
28$ and $A \le 208$.  There are 1110 known masses in this region
compared to 1654 in the region $Z,N \ge 8$ used in our standard
adjustment. Thus, about one third of all known masses are excluded,
with nuclei removed from both ends of the region of adjustment.  We
then apply the model with these constants
to the calculation of all known masses in our standard
region and compare the results to our standard model in
fig.~\ref{comp2adj}.  The error for the known nuclei is now 0.745
MeV\@, compared to 0.669 MeV with our standard model
adjusted to all known nuclei. Although there is a noticeable increase
of the error in the regions that were not included in the adjustment,
an inspection of fig.~\ref{comp2adj} indicates that the
increased error in the heavy region is not due to a systematic
divergence of the mean error, but rather to a somewhat larger scatter in
the error.

In our standard model the mass excesses of  $^{272}$110 and
$^{288}$110  are 133.82~MeV and 165.68~MeV\@, respectively. In our
restricted adjustment we obtain 133.65~MeV and 166.79~MeV\@,
respectively.  Thus, although $^{288}$110 is 80 units in $A\/$ away from
the last nucleus included in the restricted adjustment, the mass
obtained in this numerical experiment is only about 1 MeV different
from that obtained in the calculation whose constants were adjusted to
nuclei up to 50 units in $A\/$ closer to the superheavy region.  Since
our standard calculation is adjusted so much closer to the superheavy
region than is the numerical experiment, we feel that it should be
accurate to about an MeV in the superheavy region.

In fig.~\ref{actmic} we show calculated microscopic corrections
for heavy nuclei, with nuclei that are {\it calculated}
to be $\beta$-stable shown as solid squares. The region of known nuclei is
bordered by a thin solid line. The proton and neutron
drip lines, where the corresponding separation energies are zero,
are shown by thick solid lines located near the left and right edges of
the shaded region, respectively.
Minima in contour diagrams of calculated microscopic corrections are
usually associated with pairs of magic neutron and proton numbers.
Thus, in the lower-left-hand corner of the diagram we see a minimum
below $-10$ MeV\@, corresponding to the doubly magic nucleus
$^{208}_{\phantom{0}82}$Pb$_{126}$. In the upper-right-hand corner of
the figure is another minimum at proton number $Z=115$ and neutron
number $N=179$, at an energy of $-9.44$ MeV\@. At $Z=114$, $N=179$
the energy is almost the same.  This minimum is located in the region
of superheavy elements.
An interesting feature of the contour
diagram is that there is a peninsula of stability extending
from the superheavy island toward the region of known heavy elements.
On this peninsula there is a ``rock'' of increased stability centered
at $Z=109$, $N=163$.

The three heaviest known elements $_{107}$Ns, $_{108}$Hs, and $_{109}$Mt
were all identified from their $\alpha$-decay chains
 [12--14], which limited
their stability.
The single most important quantity determining the $\alpha$-decay
half-life is the $Q$ value of the decay.  In the heavy-element region
an uncertainty of 1 MeV in the $Q$ value  corresponds to uncertainties
of $10^{\pm 5}$ and $10^{\pm 3}$ for $Q_{\alpha}\approx 7$ MeV and
$Q_{\alpha}\approx 9$ MeV\@, respectively~[15].

In 1989 M\"{u}nzenberg et al. [16] compared $Q$
values for $\alpha$ decay along the $N=154$ and $N=155$ isotonic lines
to predictions of the 1988 FRLDM [17].  In
fig.~\ref{qcomp} we make a similar comparison of measured data to
predictions of the current FRDM [2]. These results based
on the current FRDM show a much improved agreement with the measured
values relative to the comparison with an older mass model in 1989 by
M\"{u}nzenberg et al.~[16].

We have estimated $\alpha$-decay half-lives  $T_{\alpha}$
corresponding to our calculated $Q_{\alpha}$ values by
use of the Viola-Seaborg systematics [18]
with parameter values
that were determined in an adjustment that included new data for
even-even nuclei [19].
The nucleus $^{272}$110 has a calculated
$\alpha$-decay half-life of about 70~ms.  The nuclei $^{288}$110 and
$^{290}$110 in the center of the superheavy island have calculated
$\alpha$-decay half-lives of 4 y and 1565 y, respectively, which, if
accurate, rules out the possibility
that superheavy elements occur in nature.

Applications of our model to the calculation of $\beta$-decay half-lives
and $\beta$-delayed neutron emission is discussed elsewhere
[20,21].

\section{Fission properties}

For a long time experimental studies of spontaneous-fission
properties showed gradual, predictable changes of such properties as
spontaneous-fission half-lives and mass and kinetic-energy
distributions as the region of known nuclei above uranium expanded.
However, in the 1970's evidence started to accumulate that
there were rapid changes in fission properties in the heavy-fermium
region. The first observation of the onset of symmetric fission at the
end of the periodic system was a study [22] of $^{257}$Fm
fission. For $^{258}$Fm the changes are even more dramatic. Fission
becomes symmetric with a very narrow mass distribution, the kinetic
energy of the fragments is about 35 MeV higher than in the asymmetric
fission of $^{256}$Fm, and the spontaneous-fission half-life is 0.38 ms,
compared to 2.86 h for $^{256}$Fm.  The fission-fragment mass
distributions and kinetic-energy distributions of $^{258}$Fm and four
other heavy nuclei are shown in fig.~\ref{hulet}, taken from
ref.~[23].
An important feature of some of the
kinetic-energy distributions  is that the shape is not Gaussian.
Instead, some of the distributions are best described as a sum of two
Gaussians.  For $^{258}$Fm, for example, the kinetic-energy
distribution can be represented by two Gaussians centered at about 200
and 235 MeV\@. This type of fission is referred to as {\it bimodal\/}
fission.

It has been proposed that the rapid change in half-life when going from
$^{256}$Fm to $^{258}$Fm is due to the disappearance of the second
saddle in the barrier below the ground-state energy. Fission through
only one barrier, the first, gives very good agreement with the
observed short half-life of $^{258}$Fm [24,25].
However, one may ask if and how the
spontaneous-fission half-life is connected to
the change in other fission properties at this transition point,
such as the change to symmetric fission and high kinetic
energies.  We show that the old interpretation that the barrier of
$^{258}$Fm has disappeared below the ground state is inconsistent with
results from the present calculation and propose a new mechanism for
the short half-life.

Although theoretical considerations had far earlier led to suggestions
of several fission paths in the potential-energy surface, theoretical
spontaneous-fission half-life calculations until rather recently
considered only shape parameterizations that allowed for the
conventional valley
[6,26--30].
Early calculations that showed, to some extent, the influence of
fragment shells at a relatively early stage of the fission process,
before scission, appeared in the early-to-mid-1970's
[31--33].

The first calculation that showed pronounced multi-valley structure
{\it and\/} predicted the corresponding spontaneous-fission half-lives
was performed in refs.~[34,35].  An improved model
that also included  odd nuclei was presented somewhat
later [36].
We show results from these calculations in figs.~\ref{shaps}--\ref{fhalf},
in units where the radius $R_0$ of the spherical nucleus is unity.
These results [34,36] showed
that some of the good agreement between calculated spontaneous-fission
half-lives and measured values obtained in earlier
calculations [25,27] for nuclei close to $^{258}$Fm was
fortuitous.

The high-kinetic-energy fragments in heavy Fm fission were
thought to correspond to fission through a scission configuration
of two touching spherical fragments, and low-kinetic-energy fission
was interpreted as fission through a scission configuration of
two elongated fragments. Figure~\ref{shaps} shows a set of shapes
that leads from a deformed ground state to both these scission configurations,
and fig.~\ref{fm258} shows the corresponding calculated potential-energy
surface. The three paths are discussed in the caption
to fig.~\ref{fm258}

In the shaded region of fig.~\ref{fm258}
we have investigated the effect of a third
mass-asymmetric deformation. The resulting most favorable shapes are
shown in fig.~\ref{shapa}, with the potential energy corresponding
to these shapes shown in fig.~\ref{cona258}. The saddle along
the long-dashed switchback path has been lowered by mass-asymmetry,
but the saddle leading to two touching spherical fragments is not
lowered by mass asymmetry. The reason that this saddle appears somewhat higher
in fig.~\ref{fm258} than in fig.~\ref{cona258} is due to interpolation
difficulties in a region of rapidly changing energy in the former figure.

Finally we present in fig.~\ref{fhalf}
calculated and measured spontaneous-fission
half-lives for some heavy elements of
interest.
Spontaneous-fission half-lives
are related to an integral along the fission path
of the product of an inertia function and the barrier along the fission
path.
Because the barrier in the valley leading to
two touching spheres is calculated to be above the ground-state energy
for $^{258}$Fm the mechanism of the short half-life is {\it not\/}
the absence of a second peak in the barrier. Instead it is a very low
inertia associated with fission in the new valley. No truly reliable
microscopic calculation of the  inertia along different fission
paths exists today, but the level structure in the new valley
suggests a very low inertia for fission along this path.

\section{Summary}

We conclude by summarizing some
important results on the
stability of the heaviest elements presented here:
\begin{itemize}
\item The inclusion of Coulomb-redistribution effects in the mass model
lowers the calculated mass for $^{272}$110 by about 3 MeV\@.
\item The superheavy island is now predicted to
be centered around
$^{288}$110 and $^{290}$110.
\item The calculated $\alpha$-decay half-lives of $^{272}$110, $^{288}$110,
and $^{290}$110 are 70 ms, 4 y, and 1565~y, respectively.
\item Relative to earlier results, we obtain shorter
spontaneous-fission half-lives in the superheavy region. For nuclei in
the vicinity of $^{272}$110 a ``ballpark'' value is 1 ms.
Thus, some spontaneous-fission half-lives
may be comparable to $\alpha$-decay
half-lives.
\item Spontaneous-fission half-lives may be significantly different from
the ``ballpark'' value of 1 ms for two reasons. One is the
general uncertainty of the calculations. Another is
that for odd systems specialization energies can
lead to huge increases in spontaneous-fission half-lives,
with up to 10 orders of
magnitude possible.
\end{itemize}

More extensive discussions of the results presented here may
be found in a series of recent publications
[1,2,10,11,20,34,36,37].

This work was supported by the U.\ S.\ Department of Energy.

\begin{small}

\end{small}

\begin{thebibliography}{10}

\bibitem{moller93:f}
P.\ M{\"{o}}ller and J.\ R.\ Nix, J.\ Phys.\ G: Nucl.\ Part.\ Phys.\ (1993) to
  be published.

\bibitem{moller93:c}
P.\ M{\"{o}}ller, J.\ R.\ Nix, W.\ D.\ Myers, and W.\ J.\ Swiatecki, {Atomic
  Data Nucl.\ Data Tables} (1993) to be published.

\bibitem{moller81:a}
P.\ {M\"{o}ller} and J.\ R.\ Nix, Nucl.\ Phys.\ {\bf A361} (1981) 117.

\bibitem{moller81:b}
P.\ {M\"{o}ller} and J.\ R.\ Nix, {Atomic Data Nucl.\ Data Tables} {\bf 26}
  (1981) 165.

\bibitem{bolsterli72:a}
M.\ Bolsterli, E.\ O.\ Fiset, J.\ R.\ Nix, and J.\ L.\ Norton, Phys.\ Rev.\
  {\bf C5} (1972) 1050.

\bibitem{moller74:b}
P.\ {M\"{o}ller} and J.\ R.\ Nix, Nucl.\ Phys.\ {\bf A229} (1974) 269.

\bibitem{krappe79:a}
H.\ J.\ Krappe, J.\ R.\ Nix, and A.\ J.\ Sierk, Phys.\ Rev.\ {\bf C20} (1979)
  992.

\bibitem{moller84:a}
P.\ {M\"{o}ller}, W.\ D.\ Myers, W.\ J.\ Swiatecki, and J. Treiner, Proc.\ 7th
  Int.\ Conf.\ on nuclear masses and fundamental constants, Darmstadt-Seeheim,
  1984 (Lehrdruckerei, Darmstadt, 1984) p.\ 457.

\bibitem{moller88:c}
P.\ M{\"{o}}ller, W.\ D.\ Myers, W.\ J.\ Swiatecki, and J.\ Treiner, {Atomic
  Data Nucl.\ Data Tables} {\bf 39} (1988) 225.

\bibitem{moller92:c}
P.\ M{\"{o}}ller and J.\ R.\ Nix, Nucl.\ Phys.\ {\bf A536} (1992) 20.

\bibitem{moller92:b}
P.\ M{\"{o}}ller, J.\ R.\ Nix, W.\ D.\ Myers, and W.\ J.\ Swiatecki, Nucl.\
  Phys.\ {\bf A536} (1992) 61.

\bibitem{munzenberg81:a}
G.\ {M\"{u}nzenberg}, S.\ Hofmann, F.\ P.\ He{\ss}berger, W.\ Reisdorf, K.-H.\
  Schmidt, J.~R.~H.\ Schneider, P.\ Armbruster, C.-C.\ Sahm, and B.\ Thuma, Z.\
  Phys.\ {\bf A300} (1981)~7.

\bibitem{munzenberg82:a}
G.\ {M\"{u}nzenberg}, P.\ Armbruster, F.\ P.\ He{\ss}berger, S.\ Hofmann, K.\
  Poppensieker, W.\ Reisdorf, J.\ R.\ H.\ Schneider, W.\ F.\ W.\ Schneider,
  K.-H.\ Schmidt, C.-C.\ Sahm, and D.\ Vermeulen, Z.\ Phys.\ {\bf A309} (1982)
  89.

\bibitem{munzenberg84:a}
G.\ {M\"{u}nzenberg}, P.\ Armbruster, H.\ Folger, F.\ P. He{\ss}berger, S.\
  Hofmann, J.\ Keller, K.\ Poppensieker, W.\ Reisdorf, K.-H.\ Schmidt, H.\ J.\
  {Sch\"{o}tt}, M.\ E.\ Leino, and R.\ Hingmann, Z.\ Phys.\ {\bf A317} (1984)
  235.

\bibitem{fiset72:a}
E.\ O.\ Fiset and J.\ R.\ Nix, Nucl.\ Phys.\ {\bf A193} (1972) 647.

\bibitem{munzenberg89:a}
G.\ {M\"{u}nzenberg}, P.\ Armbruster, S.\ Hofmann, F.\ P.\ He{\ss}berger, H.\
  Folger, J.\ G.\ Keller, V.\ Ninov, K.\ Poppensieker, A.\ B.\ Quint, W.\
  Reisdorf, K.-H.\ Schmidt, J.\ R.\ H.\ Schneider, H.\ J.\ Sch{\"{o}tt}, K.\
  S{\"{u}}mmerer, I.\ Zychor, M.\ E.\ Leino, D.\ Ackermann, U.\ Gollerthan, E.\
  Hanelt, W.\ Morawek, D.\ Vermeulen, Y.\ Fujita, and T.\ Schwab, Z.\ Phys.\
  {\bf A333} (1989) 163.

\bibitem{moller88:b}
P.\ M{\"{o}}ller and J.\ R.\ Nix, {Atomic Data Nucl.\ Data Tables} {\bf 39}
  (1988) 213.

\bibitem{viola66:a}
V.\ E.\ Viola, Jr.\ and G.\ T.\ Seaborg, J.\ Inorg.\ Nucl.\ Chem.\ {\bf 28}
  (1966) 741.

\bibitem{sobiczewski89:a}
A.\ Sobiczewski, Z.\ Patyk, and S.\ \v{C}wiok, Phys.\ Lett.\ {\bf B224} (1989)
  1.

\bibitem{moller90:a}
P.\ M{\"{o}}ller and J.\ Randrup, Nucl.\ Phys.\ {\bf A514} (1990) 1.

\bibitem{moller93:g}
K.-L.\ Kratz, J.-P.\ Bitouzet, F.-K.\ Tielemann, P.\ M{\"{o}}ller, and B.\
  Pfeiffer, Ap.\ J.\ {\bf 403} (1993) 216.

\bibitem{balagna71:a}
J.\ P.\ Balagna, G.\ P.\ Ford, D.\ C.\ Hoffman, and J.\ D.\ Knight, Phys.\
  Rev.\ Lett.\ {\bf 26} (1971) 145.

\bibitem{hulet86:a}
E.\ K.\ Hulet, J.\ F.\ Wild, R.\ J.\ Dougan, R.\ W.\ Lougheed, J.\ H.\ Landrum,
  A.\ D.\ Dougan, {M.\ Sch\"{a}del}, R.\ L.\ Hahn, P.\ A.\ Baisden, C.\ M.\
  Henderson, R.\ J.\ Dupzyk, K.\ {S\"{u}mmerer}, and G.\ R.\ Bethune, Phys.\
  Rev.\ Lett.\ {\bf 56} (1986) 313.

\bibitem{randrup73:a}
J.\ Randrup, C.\ F.\ Tsang, P.\ {M\"{o}ller}, S.\ G.\ Nilsson, and S.\ E.\
  Larsson, Nucl.\ Phys.\ {\bf A217} (1973) 221.

\bibitem{randrup76:a}
J.\ Randrup, S.\ E.\ Larsson, P.\ {M\"{o}ller}, S.\ G.\ Nilsson, K.\ Pomorski,
  and A.\ Sobiczewski, Phys.\ Rev.\ {\bf C13} (1976) 229.

\bibitem{moller72:a}
P.\ {M\"{o}ller}, Nucl.\ Phys.\ {\bf A192} (1972) 529.

\bibitem{baran81:a}
A.\ Baran, K.\ Pomorski, A.\ {\L}ukasiak, and A.\ Sobiczewski, Nucl.\ Phys.\
  {\bf A361} (1981) 83.

\bibitem{leander84:a}
G.\ A.\ Leander, P.\ {M\"{o}ller}, J.\ R.\ Nix, and W.\ M.\ Howard, Proc.\ 7th
  Int.\ Conf.\ on nuclear masses and fundamental constants (AMCO-7),
  Darmstadt-Seeheim, 1984 (Lehrdruckerei, Darmstadt, 1984) p.\ 466.

\bibitem{boning86:a}
K.\ B{\"{o}}ning, Z.\ Patyk, A.\ Sobiczewski, and S.\ {\v{C}}wiok, Z.\ Phys.\
  {\bf A325} (1986) 479.

\bibitem{staszczak89:a}
A.\ Staszczak, S.\ Pilat, and K.\ Pomorski, Nucl.\ Phys.\ {\bf A504} (1989)
  589.

\bibitem{mosel71:a}
U.\ Mosel and H.\ W.\ Schmitt, Phys.\ Rev.\ {\bf C4} (1971) 2185.

\bibitem{sandulescu76:a}
A.\ S{\u{a}}ndulescu, R.\ K.\ Gupta, W.\ Scheid, and W.\ Greiner, Phys.\ Lett.\
  {\bf 60B} (1976) 225.

\bibitem{moller77:a}
P.\ {M\"{o}ller} and J.\ R.\ Nix, Nucl.\ Phys.\ {\bf A281} (1977) 354.

\bibitem{moller87:c}
P.\ M{\"{o}}ller, J.\ R.\ Nix, and W.\ J.\ Swiatecki, Nucl.\ Phys.\ {\bf A469}
  (1987) 1.

\bibitem{moller87:d}
P.\ M{\"{o}}ller, J.\ R.\ Nix, and W.\ J.\ Swiatecki, Proc.\ Int.\
  School-Seminar on heavy ion physics, Dubna, USSR, 1986, JINR Report
  JINR-D7-87-68 (1987) p.\ 167.

\bibitem{moller89:a}
P.\ M{\"{o}}ller, J.\ R.\ Nix, and W.\ J.\ Swiatecki, Nucl.\ Phys.\ {\bf A492}
  (1989) 349.

\bibitem{moller93:b}
P.\ {M\"{o}ller}, J.\ R.\ Nix, K.-L.\ Kratz, A.\ W{\"{o}}hr, and F.-K.\
  Thielemann, Proc.\ 1st Symp.\ on nuclear physics in the universe, Oak Ridge,
  1992 (IOP Publishing, Bristol, 1993) to be published.

\end{thebibliography}
\end{document}